# The effects of aging of scientists on their publication and citation patterns

Yves Gingras, Vincent Larivière, Benoît Macaluso, Jean-Pierre Robitaille

Observatoire des sciences et des technologies (OST), Centre interuniversitaire de recherche sur la science et la technologie (CIRST), Université du Québec à Montréal, CP 8888, Succursale Centre-ville, Montréal, Québec, H3C 3P8

**Abstract**

The average age at which U.S. researchers get their first grant from NIH has increased from 34.3 in 1970, to 41.7 in 2004. These data raise the crucial question of the effects of aging on the scientific creativity and productivity of researchers. Those who worry about the aging of scientists usually believe that the younger they are the more creative and productive they will be. Using a large population of 13,680 university professors in Quebec, we show that, while scientific productivity rises sharply between 28 and 40, it increases at a slower pace between 41 and 50 and stabilizes afterward until retirement for the most active researchers. The average scientific impact per paper decreases linearly until 50-55 years old, but the average number of papers in highly cited journals and among highly cited papers rises continuously until retirement. Our results clearly show for the first time the natural history of the scientific productivity of scientists over their entire career and bring to light the fact that researchers over 55 still contribute significantly to the scientific community by producing high impact papers.

**Introduction**

A recent study by the National Institutes of Health (NIH) [1] revealed that the average age at which U.S. researchers get their first grant from that agency has increased significantly since the beginning of the 1970s. While researchers with PhDs received their first principal investigator (PI) grant at the average age of 34.3 in 1970, this figure raised to 41.7 in 2004. This increase is also observed for PIs with MDs (from 36.7 to 43.3) as well as for those having both a MD and a PhD (from 39.3 to 43.2).

Moreover, depending on the models used, it is expected that the age of new PIs could rise to 48.2 or even 54.3 in 2016. The same NIH data [2] also show that the average age of newly appointed professors in medical schools has increased from 34-36 to 37.5-40 between 1980 and 2004—depending on the diploma (MD, PhD or both). This trend is not specific to the U.S. A recent study by the Association of Universities and Colleges of Canada (AUCC) [3] showed that the average age of Canadian university professors increased from 42 years old to 49 between 1976 and 1998 and has been stable since.

These data raise the crucial question of the effects of aging on the scientific creativity and productivity of researchers. Those who worry about the aging of scientists usually believe that the younger they are the more creative and productive they will be. A better empirical knowledge of the evolution over time of productivity and creativity could thus provide important input for science policy decisions. Sociologists of science have looked into this question over the past thirty years, generally using small samples of researchers [4-9]. Contrary to expectations [10], it was found that age was not negatively correlated with productivity or creativity. This had also been observed by Dennis [11] and Adams [12] in the 40s and 50s. These findings [4-9] were coherent with Robert K. Merton's sociology of science, which suggested that age was a component of the stratification system of science: with age, scientists escalate the hierarchy of the scientific community and increase their productivity, impact and rewards [13]. In other words, the scientific community could be seen as a *gerontocracy*. Likewise, and more recently, Wray [14-15] found that it was not young scientists, but middle aged scientists, who were responsible for a disproportionate number of significant discoveries. On the other hand, recent research [16-18] still shows that young researchers (measured by either chronological or professional age) are more productive and creative than older ones given, among other things, that they have a fresh look at scientific problems [19]. This paper revisits these diverging claims and measures the effects of aging on their research productivity, scientific impact and referencing practices.

**Methods**

Using the population of Quebec's university professors and university-affiliated researchers (n=13,630), we have constructed a database containing 6,388 professors and researchers who have published at least one paper over the 8-year period (2000-2007). The average age of our population (50.4 in 2006) is similar to that of Canadian professors (49 in 2006). In order to compile meaningful statistics for each age, data is limited to professors aged between 28 and 70 years old (n≥100 university professors for each age). It must be noted that this study is cross-sectional. It does not follow the career of given individuals over time but, rather, measures differences in productivity, scientific impact and referencing patterns of professors of different age for the period 2000-2007.

All indicators in this paper are constructed using bibliometric data from Thomson Reuters' Science Citation Index Expanded™ (SCIE), Social Sciences Citation Index™ (SSCI) and Arts and Humanities Citation Index™ [20], which yearly cover the 9,000 most cited and most important journals in all fields of the natural sciences, medicine, social sciences and humanities [21]. This database list several types of scientific documents but, as usual in bibliometric studies, we limit our analysis to articles, research notes, and review articles, which are the main source of original publications [22]. Using, on the one hand, the surname and initials of professors and, on the other hand, the surname and initials of authors of scientific articles indexed by Thomson Reuters, a database of 115,342 articles authored by these professors and their namesakes was created. When papers were written in collaboration, we attributed one paper to each of the co-authors. In order to remove the papers authored by namesakes, each article has been manually validated [23]. This time-consuming but essential step reduced the number of distinct papers by 46% to 61,857.

**Results**

Figure 1.A presents the evolution of the average annual number of papers per professor, with at least one paper over the period 2000-2007, using "active" professors and "all" professors as denominators. The active professors include those who published at least one paper at that given age while the all professors includes those who have that age irrespective of the fact that they have published at that age or not (see Figure 2.A and 2.B). Both curves show that, between 28 and 40, professors sharply increase their scientific production. Then, between 41 and 50, their scientific production still rises, though at a slower pace. However, when scientists attain fifty years old, their productivity stabilizes for the rest of their career at, roughly, 3 papers per year (for active professors) or slowly decreases (for all professors). Comparing all professors and active professors clearly show that the latter, being more productive, continue to be highly productive at a later age. Of course, only a truly longitudinal analysis following the career of a cohort of scientists during decades could show if the older scientists who keep being highly productive after 60, say, were the same as those who were productive at a younger age. Our data nonetheless clearly show that active professors' scientific productivity attain its maximum during their fifties and tend to stay at that level until retirement. The decline observed for the all professors-curve can be explained by the fact that after 50, a growing fraction of professors is less active in research or has retired and stopped publishing.

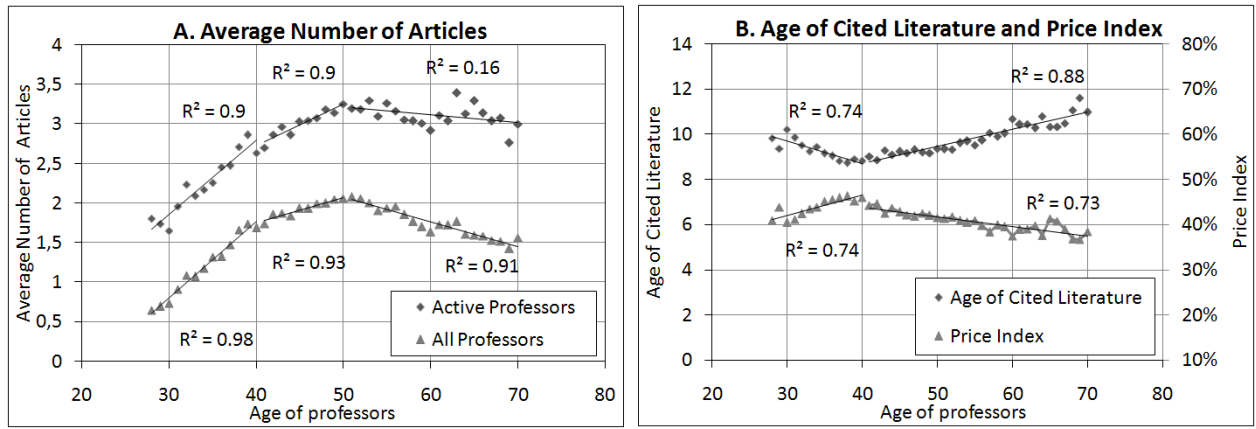

figure 1. As a function of the age of professors A) Average annual number of articles for professors who publish (Active) and for all professors B) Age of cited literature and Price Index (100 years citation window).

Figure 1.B shows the relation between the age of scientists and the average age of the cited literature. The younger the literature cited by a researcher is, the more he or she can be considered to be at the *forefront* of scientific research [24-25]. It is striking that, again, something happens when one gets 40 years old. From 28 to 40, professors rely on, and refer to, an increasingly younger body of literature. Starting at age 41, however, the literature cited ages with the author and gets older and older as time passes. This had been hypothesised by Zuckerman and Merton [26] and confirmed by Barnett and Fink [27]. Another way to look at this phenomenon is to compute the average *Price Index*—the percentage of cited references that are 5 years or younger [24]—for each age. Unsurprisingly, the same pattern is observed: professors between 28 and 40 have an increasingly high *Price Index*, which steadily falls afterward until retirement. This indicator strongly suggests that the older professors are, the more distant they are from the *most recent* (forefront) scientific research. These trends suggest a simple model of behaviour of the scientist: as a young researcher rises along the productivity curve, he/she first accumulates a basic set of references in his/her field and adds to it the most recent

papers as they appear, until he/she is about 40. After that time, a scientist tends to stick to a basic set of references and stops following as actively the growing number of recent publications.

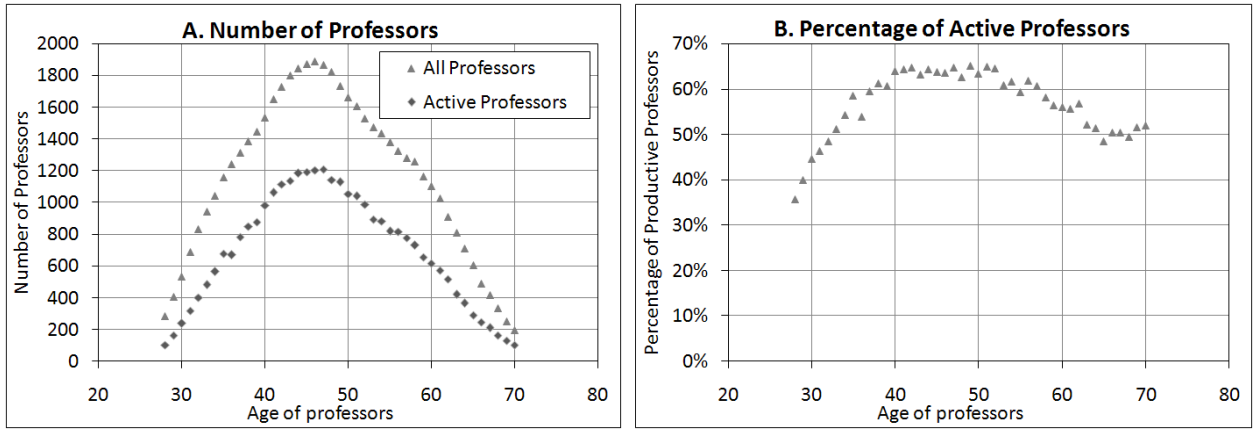

Figure 2. As a function of the age of professors A) Number of professors (all professors and active professors) B) Percentage of active professors

But does this turning point at 40 years of age also affect the scientific impact of research? To answer this question, we have calculated four different indicators: 1) the average of relative impact factor (ARIF) of the journal in which papers are published, 2) the citations received by the papers over a 3-year period following publication year (excluding self-citations) (ARC), 3) the proportion of papers in the 10% of journals with the highest (field normalized) impact factors and 4) the proportion of papers in the top 10% most cited papers (field normalized). In the calculation of the impact factors, the asymmetry between the numerator and the denominator has been corrected [28]. Also, in order to take into account the fact that publication and citation practices vary according to fields, these measures are normalized by the world average for each subfield [29-30]. When ARIF and ARC measures are above one, they are above the world average in their respective subfield, and vice-versa. As shown in Figure 3, all impact measures show a sharp decline between 28 and 50-55. However, surprisingly, impact measures subsequently rises until 70.

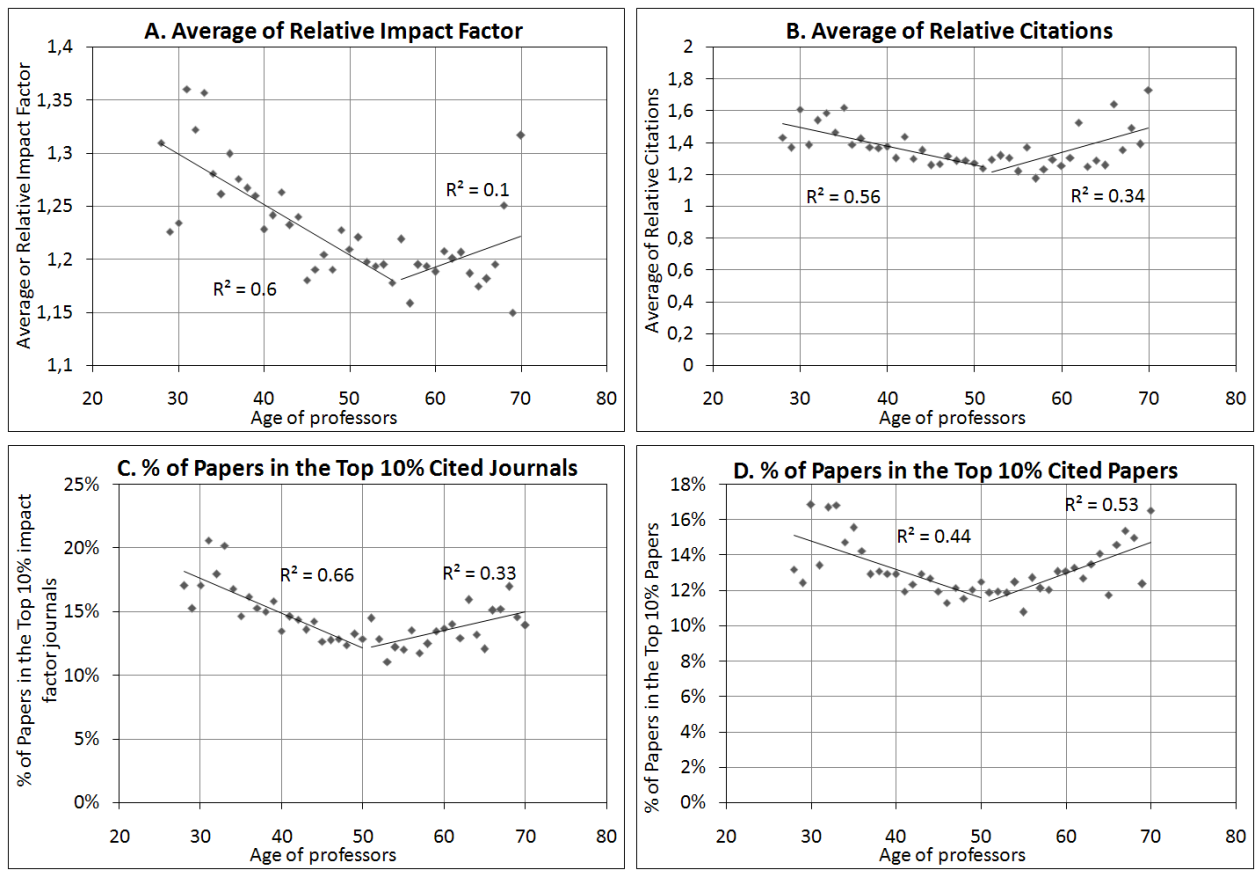

Figure 3. As a function of the age of professors A) Average of relative impact factor (ARIF) B) Average of relative citations (ARC) C) Percentage of papers in the top 10% journals D) Percentage of papers in the top 10% papers.

Part of the explanation for the fact that the average scientific impact decreases between 28 and 50 and increases afterwards while productivity is rather stable for active professors can be found in Figure 4. It presents the average number of papers per professor in the top 10% journals impact factor and top 10% most cited papers, using first *all professors* (A,B) and then only *active professors* (C,D). Figures A and B clearly show that there is a significant increase in the average number of papers in the top journals/papers for researchers aged between 28 and 40, which is rather normal given that there is also an increase in the annual number of papers for this age group. These numbers stabilize afterwards as professors publish less high impact papers after age 40; again, part of this decline can be explained by retirements. However, when only active researchers are considered as the denominator (C,D), the average number of papers in the top 10% high impact continues to rise steadily until 70. This suggests that older researchers do not publish a lower number of *high impact* papers, but rather *dilute* these high impact papers with a large number of lower impact papers, resulting in a decrease of their *average* impact. And given that they publish less after 50 (Figure 1A) and concentrate of high impact papers (Figure 4), their average impact starts to rise again (Figure 3). In sum, researchers who continue being active in research steadily increase their number of high impact papers throughout their whole career.

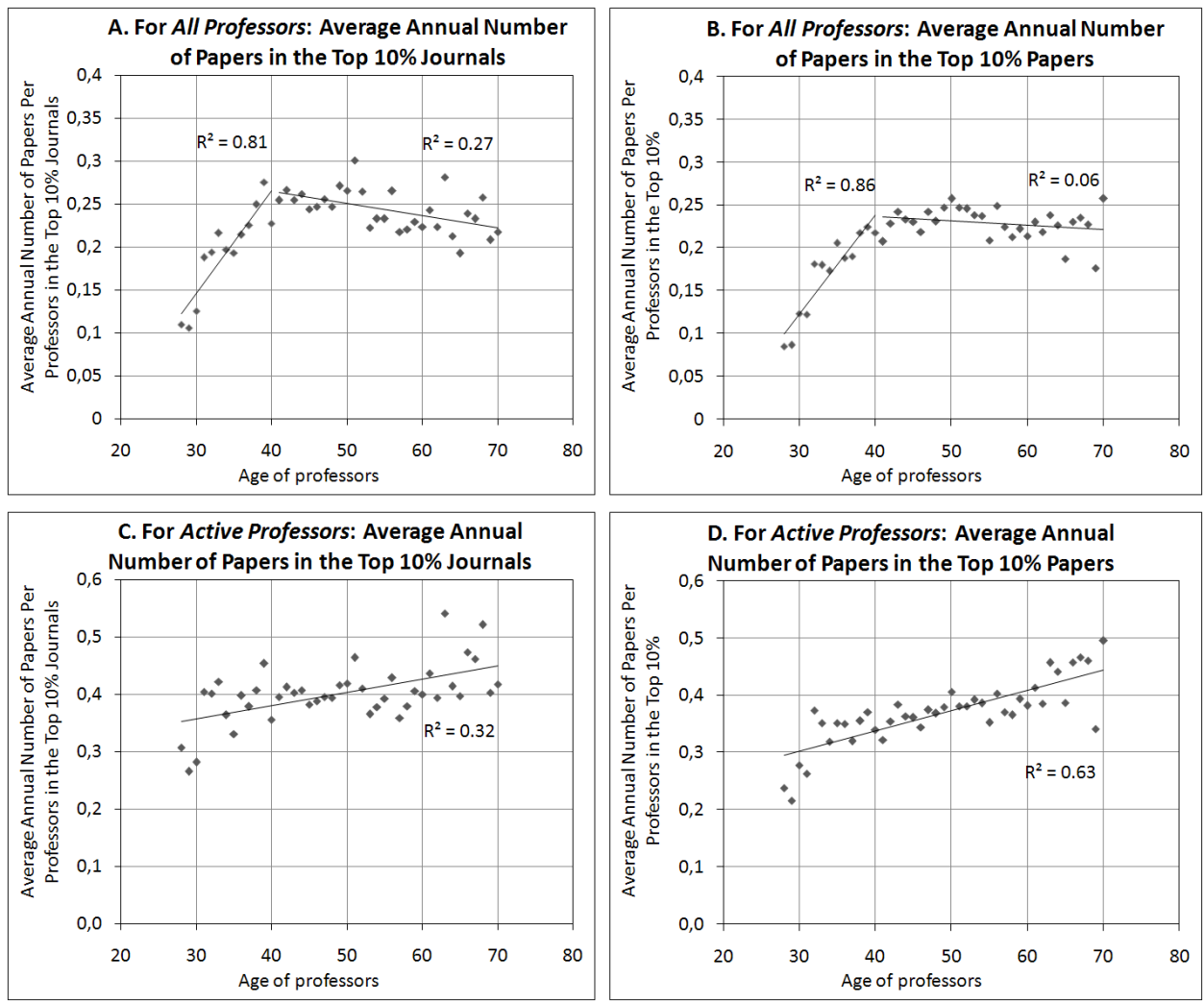

Figure 4. As a function of the age of professors A) Average annual number of papers per professor in the top 10% journals, using *all professors* as the denominator B) Average annual number of papers per professor in the top 10% papers, using *all professors* as the denominator C) Average annual number of papers per professor in the top 10% journals, using *active professors* as the denominator D) Average annual number of papers per professor in the top 10% papers, using *active professors* as the denominator.

**Discussion and Conclusion**

All the results obtained are based on a large population of researchers and clearly indentify a turning point at 40 years of age, a result consistent with other studies [31]. However, they also find another turning point—between 50 and 55 years old—where researchers are the most active but where their average scientific impact is at its lowest. Moreover, these results challenge the common belief that an aging community has less scientific impact than a younger one. Indeed, the annual production of high impact papers increases linearly with age and the average scientific impact of papers starts to rise again after 50 years old. There is also a negative relationship between productivity and impact; the years where researchers are the most active being also the years where, on average, researchers publish less impact papers.

Our analysis being cross-sectional and not longitudinal, it can contain a cohort effect which favours the younger researchers recently hired in a more competitive environment than was the case for older professors hired in the 1960s and 1970s and maybe less socialized toward high productivity. Nonetheless, the policy implications of our results are significant and show that things are more complex than expected as productivity or creativity is not a simple function of age. As discussed by Merton and Zuckerman [6, 13, 26], aging also encodes social and institutional aspects that affect the productivity of researchers. But if the turning point of 40 is also stable in a truly longitudinal sense, or in similar cohorts in other countries, providing better funding opportunities to younger researchers would at least give them more lead time to build a strong productivity before settling into a plateau.

**Acknowledgements**

This research was made possible by research contracts with Québec's Ministère de la recherche, du développement économique, de l'innovation et de l'exportation, Fonds de la recherche en santé du Québec, Fonds québécois de la recherche sur la société et la culture and the Fonds québécois de la recherche sur la nature et les technologies. We thank Stevan Harnad for comments on an earlier version.